\documentstyle[12pt]{article}
\newcommand{\es}{\\[2mm]}

\makeatother

\newcommand{\journal}[4]{{\em #1~}{\bf #2}\,(19#3)\,#4;}

\newcommand{\pr}{\journal {Phys. Rev.}}

\newcommand{\jmp}{\journal {J. Math. Phys.}}

\newcommand{\cmp}{\journal {Comm. Math. Phys.}}

\newcommand{\zp}{\journal {Z. Phys.}}
\newcommand{\np}{\journal {Nucl. Phys.}}
\newcommand{\pl}{\journal {Phys. Lett.}}
\newcommand{\mpl}{\journal {Mod. Phys. Lett.}}

\newcommand{\annp}{\journal {Ann. Phys. (N.Y.)}}



\def\Lp{\displaystyle{\biggl(}}
\def\Rp{\displaystyle{\biggr)}}
\def\LP{\displaystyle{\Biggl(}}
 
\def\RP{\displaystyle{\Biggr)}}
\newcommand{\lp}{\left(}\newcommand{\rp}{\right)}

\renewcommand{\a}{\alpha}
\renewcommand{\b}{\beta}
\renewcommand{\d}{\delta}
\newcommand{\e}{\varepsilon}
\newcommand{\f}{\phi}

\renewcommand{\l}{\lambda} 
\newcommand{\m}{\mu}
\newcommand{\n}{\nu}

\newcommand{\s}{\sigma} \renewcommand{\S}{\Sigma}

\newcommand{\complex}{{\kern .1em {\raise .47ex
\hbox {$\scriptscriptstyle |$}}
    \kern -.4em {\rm C}}}
\newcommand{\real}{{{\rm I} \kern -.19em {\rm R}}}
\newcommand{\rational}{{\kern .1em {\raise .47ex
\hbox{$\scripscriptstyle |$}}
    \kern -.35em {\rm Q}}}
\renewcommand{\natural}{{\vrule height 1.6ex width
.05em depth 0ex \kern -.35em {\rm N}}}
\newcommand{\dint}{\displaystyle{\int}}
\newcommand{\xint}{\dint d^4 \! x \, }

\newcommand{\pa}{\partial}

\newcommand{\fud}[2]  {{\displaystyle{\frac{\delta #1}{\delta #2}}}}
\newcommand{\dfrac}[2]{{\displaystyle{\frac{#1}{#2}}}}

\newcommand{\sla}{\raise.15ex\hbox{$/$}\kern -.57em}

\newcommand{\twiddle}{\lower.9ex\rlap{$\kern -.1em\scriptstyle\sim$}}

\newcommand{\equ}[1]{(\ref{#1})}

\newcommand{\eq}{\begin{equation}}
\newcommand{\eqn}[1]{\label{#1}\end{equation}}
\newcommand{\eea}{\end{eqnarray}}
\newcommand{\eqa}{\begin{eqnarray}}
\newcommand{\eqan}[1]{\label{#1}\end{eqnarray}}
\newcommand{\ba}{\begin{array}}
\newcommand{\ea}{\end{array}}
\newcommand{\eqac}{\begin{equation}\begin{array}{rcl}}
\newcommand{\eqacn}[1]{\end{array}\label{#1}\end{equation}}

\newcommand{\at}{{\~a}}

\newcommand{\ooo}{{\'o}}

\newcommand{\iii}{{\'\i}}

\begin{document}
\begin{titlepage}

\begin{flushright}

CBPF-NF-041/97\\
hep-th/xxxxxxx\\

\end{flushright}
\vfill

\begin{center}
{\Large{\bf A No-Go Theorem for the Nonabelian  Topological Mass Mechanism in 
Four Dimensions }}

\end{center}
\vfill

\begin{center}
{\large
M. Henneaux$^{a,b}$,
V.E.R. Lemes$^c$, C.A.G. Sasaki$^c$,
S.P. Sorella$^d$,  \\
O.S. Ventura$^c$ 
and L.C.Q. Vilar$^c$}
\end{center}
\vfill

\begin{center}{\sl
$^a$ Facult\'e des Sciences, Universit\'e Libre de
Bruxelles,\\
Campus Plaine C.P. 231, B--1050 Bruxelles, Belgium\\[1.5ex]

$^b$ Centro de Estudios Cient\'\i ficos de Santiago,\\
Casilla 16443, Santiago 9, Chile\\[1.5ex]

$^c$ C.B.P.F, Centro Brasileiro de Pesquisas F{\iii}sicas,\\
Rua Xavier Sigaud 150, 22290-180 Urca, Rio de Janeiro, Brazil\\[1.5ex]

$^d$ UERJ, Universidade do Estado do Rio de Janeiro,\\
Departamento de F{\iii}sica Te{\ooo}rica, Instituto de F{\iii}sica, UERJ,\\
Rua S{\at}o Francisco Xavier, 524, 20550-013 Maracan{\at}, Rio de Janeiro, Brazil

}\end{center}
\vfill

\begin{abstract}
We prove that there is
no power-counting renormalizable nonabelian generalization of the abelian
topological mass mechanism in four dimensions.
The argument is based on the technique of
consistent deformations of the master equation
developed by G. Barnich and one of the authors.
Recent attempts involving extra fields are also 
commented upon.

\end{abstract}


\end{titlepage}

\section{Introduction}
\setcounter{equation}{0}

One of the most intriguing issues of high energy physics is the 
understanding of the mechanism which provides the masses for the 
gauge vector bosons. As it is well known the Higgs mechanism, 
although consistent
with the renormalizability and the unitarity constraints of quantum field 
theory,
relies on the existence of scalar particles, the Higgs bosons, whose 
experimental
evidence is still lacking.

Therefore any new alternative mechanism to generate masses is welcome, deserving
attention and careful analysis. In particular, the idea that the vector boson masses
could originate from a topological mechanism preserving exact gauge invariance is
rather tempting and fascinating. The example provided by the topological nonabelian
Chern-Simons term in three space-time dimensions is certainly the most celebrated
way to provide a topological mass for the Yang-Mills fields~\cite{cs}.
In four dimensions, an analogous mechanism has been 
proposed in the abelian
case~\cite{fourd}. It makes use of a two-form gauge field
$B_{\m\n}=-B_{\n\m}$ suitably coupled to the one-form gauge connection $A_{\m}$.
The action reads 
\eq
S^{ab}_{m} = \xint \LP -\dfrac{1}{4}F_{\m\n}F^{\m\n} -
\dfrac{1}{12}H_{\m\n\rho}H^{\m\n\rho} +
\dfrac{1}{2}m\e^{\m\n\rho\l}F_{\m\n}B_{\rho\l} \RP \ ,
\eqn{ab-action}
with $F_{\m\n}  = (\pa_{\m}A_{\n} - \pa_{\n}A_{\m})$ and $H_{\m\n\rho}$ the totally
antisymmetric three-form
\eq
 H_{\m\n\rho}  = \pa_{\m}B_{\n\rho} +  \pa_{\n}B_{\rho\m} + \pa_{\rho}B_{\m\n} \ ,
\eqn{3-form}
The action \equ{ab-action} is invariant under
two kinds of local transformations given  by
\eq
 \d^{g}A_{\m} = \pa_{\m}\e \ ,  \qquad \d^{g}B_{\m\n}=0 \ ,
\eqn{ab-gauge-tr}
and
\eq
 \d^{t}A_{\m} = 0 \ ,  \qquad \d^{t}B_{\m\n}= \pa_{\m} \e_{\n} - \pa_{\n} \e_{\m} \ .
\eqn{ab-tens-tr}
Equations~\equ{ab-gauge-tr} and \equ{ab-tens-tr} correspond respectively to
ordinary gauge transformations and to vector type transformations related
to the tensorial character of the field $B_{\m\n}$. 
The parameter $m$ of the topological
term $\e^{\m\n\rho\l}F_{\m\n}B_{\rho\l}$ in the expression
\equ{ab-action} has the dimension of a mass.
It is easily verified that this term gives a nonvanishing pole
for the physical two-point function, yielding thus a
gauge-invariant topological
mass~\cite{fourd}.  The two-form $B_{\mu \nu}$ is actually dual to the
scalar field that is eaten up by the gauge field in the standard Higgs
mechanism.  There is, however, no additional particle and the action~\equ{ab-action}
describes a single massive vector field without Higgs boson.

The aim of this work is to discuss to what extent the abelian model \equ{ab-action}
can be generalized to the nonabelian case. The analysis will be performed by
making use of the method of the consistent deformations developed
in \cite{def}. We will end
up, unfortunately, with a  no-go theorem stating that it is not possible
to generalize the expression \equ{ab-action} to a local, power counting
renormalizable, nonabelian action while preserving the same field content
and the same number of local symmetries. In other words, possible nonabelian
generalizations of the action \equ{ab-action} will necessarely require 
non-renormalizable couplings, as in \cite{FT}, or the
introduction of extra fields \cite{attempt}.

The paper is organized as follows. In Sect.II we briefly review 
the method of consistent deformations. Sect.III
will be devoted to the detailed proof of the aforementioned no-go theorem. 
Sect. IV discusses the assumptions of the
no-go theorem and comments on possible ways out.

\section{Consistent Deformations}
\setcounter{equation}{0}

The method that we shall follow is based on the antifield-BRST formalism
and allows to construct interaction terms by consistently deforming the
master equation\footnote{The usefulness of the deformation point of view
(but not in the general framework
of the antifield formalism, which allows off-shell open deformations
of the algebra) has been advocated 
in \cite{Julia}.
For a recent discussion emphasizing the homological aspects, see \cite{Stasheff}.}. 
Following the original work~\cite{def}, the starting point
is a given action $S_{inv}[\f^{a}]$ with local gauge symmetries
\eq
 \d_{\e}\f^{a}=R^a_{\a}\e^{\a}, \qquad \d_{\e} S_{inv}[\f^{a}] = 0\ .
\eqn{start-point}
According to  the well known antifield formalism (for a review
appropriate to the
subsequent cohomological considerations, see~\cite{ht}), 
we introduce ghost fields $C^{\a}$ and a suitable set of
antifields $\f^{\ast A}$, so that the action
\eq
S_{0}[\f^{A},\f^{\ast A}] = S_{inv}[\f^{a}] + \dint d^4x \f^{\ast}_{a}R^a_{\a}C^{\a}
                             + ....  \ ,
\eqn{m-e-sol}
is a solution of the master equation
\eq
\lp S_{0},S_{0} \rp =
\xint \fud{S_{0}}{\f^{A}} \fud{S_{0}}{\f^{\ast A}} = 0 \ ,
\eqn{master-eq}
where $\f^{A} = \lp \f^{a},C^{\a} \rp $ denote collectively all the fields and ghosts.
The BRST differential $s$ in the space of the fields
and antifields is defined through the antibraket
\eq
s \f^{A} = \lp \f^{A},S_{0} \rp = \fud{S_{0}}{\f^{\ast A}}\ , \qquad
s \f^{\ast A} = \lp \f^{\ast A},S_{0} \rp =\fud{S_{0}}{\f^{A}} \ .
\eqn{atua-master}
Let us suppose now that the action $S_{0}$ refers to a free field theory
which does not contain any coupling constant or mass parameter and let us
ask ourselves if it is possible to introduce consistent interactions for $S_{0}$, {\it i.e.}
\eq
S_{0} \rightarrow S  = S_{0} + g_{i}S_{i}
+ g_{i}g_{j}S_{ij} + ... \ ,
\eqn{interaction}
in such a way that the resulting action $S$ still satisfies the deformed master equation
\eq
 \lp S,S \rp = 0 \ .
\eqn{def-master-eq}
Of course, due to locality and power counting, we shall limit ourselves to
interactions $S_{i}$ which are integrated local polynomials in the fields
and antifields with dimensions less or equal to four. Accordingly, the
expansion parameters $g_{i}$ will be required to have nonnegative mass
dimension. Therefore they will have the meaning of coupling constants and masses.
As shown in ref.~\cite{def}, the requirement of the validity of the master
equation \equ{def-master-eq} automatically implies the existence of a
deformed action $S^{g}_{inv}[\f^{A}]$,
\eq
S^{g}_{inv}[\f^{A}]= S[\f^{A}, \f^{\ast A}=0]=S_{inv}[\f^{a}]
                    + O(g_i)  \ ,
\eqn{def-action}
which is left invariant under a consistent deformed version of the original
gauge symmetries, {\it i.e.}
\eq
    \d^{g}_{\e}S^{g}_{inv}[\f^{A}]= 0 \ ,
\eqn{def-gauge-inv}
with
\eq\ba{rl}
 \d^{g}_{\e}\f^{a} =& \fud{S}{\f^{\ast a}}[\f^{A}, \f^{\ast A}=0] \ , \qquad
  (C^{\a} \rightarrow \e^{\a}) \ , \\ \es
\d^{g}_{\e}\f^{a}= & R^a_{\a}\e^{\a} + O(g_i) \ .
\ea\eqn{expl-def-symm}
Equations \equ{def-action}-\equ{expl-def-symm} mean indeed that we have been
able to add invariant interaction terms to the original free action by
suitable deforming the gauge symmetry, the construction being done order
by order in the parameters $g_{i}$.
For instance, in the case in which as the free action $S_{0}$ one chooses the
Maxwell lagrangian, the above construction naturally leads to the nonabelian
Yang-Mills action with the well known cubic and quartic interaction terms.
Moreover, the usefulness of working with the master equation lies in the fact
that the search of possible nontrivial\footnote{According to ref~\cite{def},
an interaction  term is called  trivial  if it can obtained through a field
redefinition.}
interaction terms can be reduced to a cohomology problem for the free BRST
differential of eq.\equ{atua-master}, whose cohomology classes are eventually
known or amenable to tractable computations.
For a better understanding of
this point, let us expand the master equation \equ{def-master-eq} in powers
of the deformation parameters $g_i$:
\eq\ba{rl}
& \lp S_{0},S_{0} \rp = 0 \, ,\es \\
&  \lp S_{0},S_{i} \rp = 0 \, ,\es \\
& 2 \lp S_{0},S_{ij} \rp + \lp S_{i},S_{j} \rp = 0 \, ,\es \\
& .......
\ea\eqn{master-ordens}
The first equation is nothing but the master equation for the free theory
$S_0$, and is satisfied by hypothesis. From the second condition we see that
$S_{i}$ has to be  invariant under the action of the free BRST differential
$s \equiv \lp .,S_{0} \rp $. However, interactions of the type
$S_{i} = \lp T_{i},S_{0} \rp$, for some integrated local $T_{i}$,
have to be neglected since they are seen to correspond to pure field
redefinitions~\cite{def}. This means that the nontrivial interaction terms
which can be added to the free action $S_{0}$ in the first order approximation 
in the  parameters $g_{i}$ have to belong to the cohomology of the BRST
differential $s$. Concerning  the third equation, it is very easy to see
that it can admit a solution only if the antibracket  $\lp S_{i},S_{j} \rp$
can be written in the form of an exact cocycle, {\it i.e.} $\lp S_{i},S_{j}
\rp = \lp T_{ij},S_{0} \rp$, for some local $T_{ij}$. Otherwise, if $\lp
S_{i},S_{j} \rp$ belongs to the cohomology of $s$ we have an obstruction
whose effect is to generate constraints among the various parameters
$g_{i}$. The same conclusions hold for the higher order levels of the expansion
of the deformed master equation \equ{def-master-eq}. In other words, at each
step, the  parameters
$g_{i}$ are required  to fulfill a certain number of conditions. It may
happen that the conditions met at a certain level can be fulfilled
only if some of the $g_{i}$ 's    vanish.  This means that 
the corresponding first-order deformation is obstructed,
so that the full deformation program can be 
achieved only for a restricted susbset of the couplings.
As we shall see in the next section, this will be the case of the nonabelian
generalization we are looking for. In fact,  we shall be able to prove that
the abelian action \equ{ab-action} and the gauge symmetries of eqs. \equ{ab-gauge-tr}
and \equ{ab-tens-tr} can be consistently deformed only if $mg=0$
(where $g$ is the Yang-Mills coupling constant), implying
that there is no nonabelian, power-counting renormalizable
generalization of the topological mass term
$\e^{\m\n\rho\l}F_{\m\n}B_{\rho\l}$.

\section{The No-Go Theorem}
\setcounter{equation}{0}

Let us now apply the previous construction to the analysis
of a possible nonabelian extension of the action
\equ{ab-action}. We shall start therefore with the following
free gauge invariant action
\eq
\S_{0} = \xint \LP -\dfrac{1}{4}{F^{a}}_{\m\n}F^{a \m\n} -
\dfrac{1}{12}{H^{a}}_{\m\n\rho}H^{a \m\n\rho} \RP \, ,
\eqn{inicial}
where $F^{a}_{\m\n}$ and $H^{a \m\n\rho}$ are the abelian curvatures
\equ{3-form} for a set of $n$ $(a=1,..,n)$ gauge and tensor
fields
$A^{a}_{\m}$ and $B^{a}_{\m\n}$. It is worth recalling here
that, within the consistent deformation set up, the mass
parameter $m$ of eq.\equ{ab-action} is considered, as any
other coupling, as a deformation parameter. Accordingly, the
topological mass term $\e^{\m\n\rho\l}F_{\m\n}B_{\rho\l}$
will in fact appear as a first order consistent deformation.

Of course, the free action \equ{inicial} is invariant under
the gauge transformations
\eq
\d A^{a}_{\m}  = \pa_{\m}\e^{a} \ , \qquad
\d B^{a}_{\m\n}  = \pa_{\m} \e^{a}_{\n} - \pa_{\n} \e^{a}_{\m} \ .
\eqn{transf}
Taking into account that the trasnformation of $B^{a}_{\m\n}$
is reducible, due to the existence of the zero modes
$\d\e^{a}_{\m}=\pa_{\m}\omega^{a}$, we introduce a set of ghosts $(c^a,
\eta^{a\m},\rho^a)$, where $c^a$ and $\eta^{a\m}$ stand for the ghost
corresponding to the gauge 
transformations \equ{transf} and $\rho^a$ is a ghost for ghost accounting
for the reducibility. 
Moreover, introducing also the antifields 
$(A^{\ast \m}_{a},B^{\ast\m\n}_{a},c^{\ast}_{a},\eta^{\ast\m}_{a},
\rho^{\ast}_{a})$ and the corresponding action $\S_{ant}$ 
\eq
\S_{ant} = \xint \LP A^{\ast \m}_{a}\pa_{\m}c^{a} +
B^{\ast\m\n}_{a}\pa_{\m}\eta^{a}_{\n} +
\eta^{\ast\m}_{a}\pa_{\m}\rho^{a} \RP \ ,
\eqn{antacao}
one finds that the complete {\it free} action 
\eq
S_{0}\,  = \, \S_{0} \, + \, \S_{ant} \ ,
\eqn{tudo}
satisfies the master equation 
\eq
\lp S_{0},S_{0} \rp = 0 \ ,
\eqn{free-mast-eq}
with
\eq
\lp S_{0},S_{0} \rp =  \xint \LP 
                \fud{S_{0}}{A^{a\m}} \fud{S_{0}}{A^{\ast}_{a\m}}
   + \dfrac{1}{2} \fud{S_{0}}{B^{a\m\n}} \fud{S_{0}}{B^{\ast}_{a\m\n}}   
   + \fud{S_{0}}{c^{a}} \fud{S_{0}}{c^{\ast}_{a}} 
  + \fud{S_{0}}{\eta^{a\m}} \fud{S_{0}}{\eta^{\ast}_{a\m}}
   + \fud{S_{0}}{\rho^{a}} \fud{S_{0}}{\rho^{\ast}_{a}} \RP \ .
\eqn{expl-free-m-e}
According to Eq.\equ{atua-master}, the nilpotent BRST transformation is 
\eq\ba{ll}
 s A^{a}_{\m} = \pa_{\m}c^{a} \, ,
& s A^{\ast \m}_{a}  = \pa_{\n}F^{a \n\m} \, ,\\ \es
 s B^{a}_{\m\n}  = \pa_{\m} \eta^{a}_{\n} - \pa_{\n}
 \eta^{a}_{\m} \, ,
& s B^{\ast\m\n}_{a}  = \pa_{\rho}H^{\rho\m\n}_{a} \, ,\\ \es
 s \eta^{a}_{\m}  = \pa_{\m}\rho^{a} \, ,
& s \eta^{\ast\m}_{a}  = - \pa_{\n}B^{\ast\m\n}_{a} \, ,\\ \es
 s c^{a}  = 0 \, ,
& s c^{\ast}_{a}  = - \pa_{\m}A^{\ast \m}_{a} \, ,\\ \es
 s \rho^{a}  = 0 \, ,
& s \rho^{\ast}_{a}  = \pa_{\m}\eta^{\ast\m}_{a} \, .\\ \es
\ea\eqn{transforma}
We should remark that the antifields $(c^{\ast}_{a},\rho^{\ast}_{a})$
conjugate  to the ghosts $(c^a,\rho^a)$, although not explicitely appearing
in the expression \equ{antacao} due to the fact that $c^a$ and $\rho^a$ do
not transform under $s$ in the free abelian limit, are  needed
to allow a priori for general deformations deforming also the gauge
algebra or the reducibility functions~\cite{def}. 
Let us display, for further use, the quantum numbers of all the fields and antifields. 
\begin{table}[hbt]
\centering
\begin{tabular}{|c|c|c|c|c|c|c|c|c|c|c|} \hline
 & $A^{a}_{\m}$ & $B^{a}_{\m\n}$ & $c^{a}$ & $\eta^{a}_{\m}$ & $\rho^{a}$
 & $A^{\ast a}_{\m}$ & $B^{\ast a}_{\m\n}$ & $c^{\ast a}$
& $\eta^{\ast a}_{\m}$ & $\rho^{\ast a}$\\ \hline
$N_{g}$ & 0 & 0 & 1 & 1 & 2 & -1 & -1  & -2 & -2 & -3  \\ \hline
$dim$ & 1 & 1 & 0 & 0 & -1 & 3 & 3  & 4 & 4 & 5  \\ \hline
\end{tabular}
\caption[t1]{Ghost numbers and dimensions.}
\label{gh1-number-dim}
\end{table}

We face now the problem of characterizing the possible interactions terms $S_{i}$ 
that can be added to the action \equ{tudo} while preserving the
master equation \equ{free-mast-eq}. Following the algebraic set up of the
previous Section, the consistent interactions which can be introduced to the
first order in the deformation parameters are nontrivial solutions of the
consistency condition 
\eq  
 \lp S_{0},S_{i} \rp = 0 \ ,
\eqn{cons-cond}
with $S_{i}$ local integrated polynomials of ghost number zero and dimension
bounded by four. 

The solutions of \equ{cons-cond} for free  $p$-form
gauge fields have been studied in \cite{pf} and fall into three
categories: (i) those that do not deform the gauge symmetry;
(ii) those that deform the gauge transformations but not the
gauge algebra; and (iii) those that deform both the gauge transformations 
and the gauge algebra.  

The first category contains the gauge-invariant
functions (= functions of the curvature components and their
derivatives) as well as the Chern-Simons terms.  The only candidates allowed by
Lorentz invariance and power-counting are the kinetic
terms  $\sim F^2$, $H^2$ - which are already present in the Lagrangian - and the
Chern-Simons term $\e^{\m\n\s\l}B^{a}_{\m\n}F^{a}_{\s\l}$ (the
Chern-Simons term $\e^{\m\n\s\l} H^{a}_{\m\n\s} A^{a}_{\l}$
differs from $\e^{\m\n\s\l}B^{a}_{\m\n}F^{a}_{\s\l}$
by a total derivative and hence is not independent from it).
Consequently, there is only one independent new vertex in the first
category, which
we denote by $S_2$,
\eq
S_{2} = m_{ab} \xint \Lp \e^{\m\n\s\l}B^{a}_{\m\n}F^{b}_{\s\l} \Rp \ .
\eqn{S2}
Here, $m_{ab}$ is a mass matrix which may have zero eigenvalues, reflecting
the possibility that some gauge bosons may remain massless.

The second category involves interactions of the Noether form
$A^a_\mu j_a^\mu$ and $B^a_{\mu \nu} k_a^{\mu \nu}$, where $j_a^\mu$ are 
gauge-invariant conserved currents, while $k_a^{\mu \nu}$ are gauge-invariant
conserved antisymmetric tensors ($\partial_\mu k_a^{\mu \nu} = 0$).
There are only two types of non-trivial gauge-invariant
conserved antisymmetric tensors, which are
$F_a^{\mu \nu}$
and $\epsilon^{\mu \nu \rho \sigma} 
\overline{H}^a_\rho \overline{H}^b_\sigma$ \cite{pf}.
Here, $\overline{H}^a$ denotes the one-form dual to $H^a$.
The conserved tensors $\epsilon^{\mu \nu \rho \sigma}\overline{H}^a_\rho 
\overline{H}^b_\sigma$, which lead to the Freedman-Townsend
coupling \cite{FT}, are excluded by power-counting renormalizability
since they require coupling constants with dimensions of an inverse mass.
This leaves us with $B^a_{\mu \nu} F^{b \mu \nu}$ only.
Similary, even though there is an infinite number
of gauge invariant conserved currents $j^a_\mu$, Lorentz invariance and
power-counting renormalizability exclude all couplings of the
form $A^a_\mu j_a^\mu$ except $t_{ab} A^a_\mu \e^{\m\n\s\l} H^b_{\m\n\s}$
and $ t_{ab} A^a_{\m} \partial_\nu F^{b \m \n}$.
But $t_{ab} A^a_\mu \e^{\m\n\s\l} H^b_{\m\n\s}$  is equivalent to the above Chern-Simons 
term, while $ t_{ab} A^a_{\m} \partial_\nu F^{b \m \n}$ vanishes on-shell and can be 
absorbed by redefinitions.
To summarize, there is accordingly only one novel coupling in the second category,
namely $\mu_{ab} B^a_{\mu \nu} F^{b \mu \nu}$, with matrix
$\mu$ having the dimensions of a mass.  This coupling is of the Chapline-Manton
type \cite{CM} and has the following BRST invariant extension,
\eq
S_{3}  = \mu_{ab} \xint \Lp \dfrac{1}{2} B^{a}_{\m\n}F^{b \m\n} +
A^{\ast a \m}\eta^{b}_{\m} + c^{\ast a}\rho^{b} \Rp \ .
\eqn{S3}
It is rather interesting
to remark that the expression \equ{S3} allows, at least in the abelian case,
for a further nontopological mass mechanism. It is indeed immediate to check
that the following abelian action
\eq\ba{ll}
S^{ab}_{\m} & = \xint \Lp -\dfrac{1}{4}( F_{\m\n} - \m B_{\m\n} )^2
+ \dfrac{1}{12}H_{\m\n\rho}H^{\m\n\rho}  \Rp \\ \es
& = \xint \Lp -\dfrac{1}{4} F_{\m\n}F^{\m\n} + \dfrac{1}{12}H_{\m\n\rho}H^{\m\n\rho}
  +\dfrac{\m}{2}B^{\m\n}F_{\m\n} - \dfrac{\m^2}{4}B_{\m\n}B^{\m\n} \Rp ,
\ea\eqn{ab-mu-action}
is left invariant by the transformations
\eq
\d A_{\m} = \pa_{\m}\e + \m \e_{\m} \ ,  \qquad
\d B_{\m\n}= \pa_{\m} \e_{\n} - \pa_{\n} \e_{\m} \ .
\eqn{mu-transf}
The new terms in \equ{ab-mu-action}
provide a nontopological mass for the two-form 
field $B_{\m \n}$. The action \equ{ab-mu-action} was actually already considered
in \cite{KR}.  Note that the transformation of the gauge connection
gets modified by an extra $\m$-dependent term which enables one to gauge it
away.  Recall also that a massive two-form in four dimensions describes massive
spin-one particles, just as a massive one-form.

Finally, the third category contains only the familiar Yang-Mills
cubic interaction vertex with dimensionless parameter $g$ \cite{pf} (second reference),
\eq  
S_{1}  = g \xint f^{a b c}\Lp -\dfrac{1}{2} F^{a}_{\m\n}
A^{b\m}A^{c\n} + A^{\ast a}_{\m}A^{b\m}c^{c} +
\dfrac{1}{2}c^{\ast a}c^{b}c^{c} \Rp \ .
\eqn{S1}
It should be stressed, in particular, that \equ{S1} {\em cannot}
be accompanied by a deformation of the type $A^a_\mu B^b_{\rho \sigma}
H^{c \mu \rho \s} f_{abc}$.  Such an interaction vertex  is the minimal coupling
to the Yang-Mills field of a charged $B$-field transforming in the adjoint.
It has the Noether form $A^a_\mu j^\mu_a$ where the current
$j^\mu_a = B^b_{\rho \sigma} H^{c \mu \rho \s} f_{abc}$ is, however,
{\em not} gauge invariant under the gauge transformations of the
two-forms.  For this reason, $A^a_\mu B^b_{\rho \sigma}
H^{c \mu \rho \s} f_{abc}$ is also not gauge-invariant, even
up to a total derivative, and does not define
an observable.  Thus, it does not lead to a consistent deformation.  The
two-forms $B^a_{\rho \sigma}$ are uncharged \cite{DeserWitten}
and can only couple to the
connection $A$ through Chern-Simons or Chapline-Manton terms
(to first order).

Having characterized the possible nontrivial consistent interaction
terms that can be added to  first order in the deformation parameters
$(g,m,\m)$, let us turn to the study of the higher order consistency conditions
stemming from the requirement of validity of the deformed master equation
to all orders.
As we have already seen, the second order consistency condition, {\i.e.} the
third equation of the system \equ{master-ordens}, can be solved only if the
antibrackets $\lp S_{i},S_{j} \rp$, with $i,j=1,2,3$, can be written as exact
BRST cocycles. 
The antibracket $\lp S_{2},S_{2} \rp$ is automatically vanishing, due to the
fact that $S_{2}$ is independent from the antifields. In addition, the
constraints which follow 
from $\lp S_{1},S_{1} \rp$ are the usual ones which 
lead to the pure Yang-Mills vertices and are satisfied by identifying the
$f^{abc}$ 's in eq.\equ{S1} with the structure constant of a Lie group.
Concerning now the antibrackets $\lp S_{2},S_{3} \rp$ and $\lp S_{3},S_{3}
\rp$, they are easily seen to be BRST trivial 
\eq\ba{rl}
(S_{2},S_{3}) & = - \dfrac{2}{3} m \m \xint
\e^{\m\n\rho\s}H^{a}_{\m\n\rho}\eta^{a}_{\s} = s \Lp  \dfrac{ m \m}{2}\xint
\e^{\m\n\rho\s}B^{a}_{\m\n}B^{a}_{ \rho\s} \Rp \, , \\ \es
(S_{3},S_{3}) & = - 2 \m^{2} \xint \pa^{\m}B^{a}_{\m\n}\eta^{a \n} =
s \Lp  \dfrac{\m^{2}}{2}\xint B^{a \m\n}B^{a}_{\m\n} \Rp \, ,
\ea\eqn{triviais}
so that they do not bring any obstruction. It remains therefore to analyse the terms  
$\lp S_{1},S_{3} \rp$ and $\lp S_{1},S_{2} \rp$, given respectively by 
\eq\ba{rl}
(S_{1},S_{3}) = - g \m \xint f^{abc}\Lp & F^{a}_{\m\n}A^{b
\m}\eta^{c \n} - \pa^{\m}(A^{b}_{\m}A^{c}_{\n})\eta^{a\n} +
(\pa^{\m}B^{a}_{\m\n})A^{b \n}c^{c} \\ \es
&  - A^{\ast a}_{\m}c^{c}\eta^{b\m} 
   - A^{\ast a}_{\m}A^{b \m}\rho^{c} - c^{\ast a}c^{b}\rho^{c} \Rp \,
    \\ \es
(S_{1},S_{2}) & = - \dfrac{2}{3} m g
\xint f^{abc}\e^{\m\n\rho\s}H^{a}_{\m\n\rho}A^{b}_{\s}c^{c} \, .
\ea\eqn{importantes}

It is not difficult to convince oneself that the above expressions
are not BRST trivial, representing thus a real obstruction to 
the deformation of the master equation. Concerning the first expression, we
indeed observe that the antifield dependent terms do not contain any space-time
derivative. On the other hand, from eqs.\equ{transforma} we see that the BRST
transformations of the antifields always introduce a space-time derivative,
implying thus that $\lp S_{1},S_{3} \rp$ cannot be cast in the form of a pure
$s$-variation. Similarly, by means of a simple counting of the space-time
derivatives appearing in the second term of eq.\equ{importantes} as well as
in the BRST transformations \equ{transforma}, we can easily verify that the
only possible candidate whose BRST variation could reproduce the second
antibracket  $\lp S_{1},S_{2} \rp$ is given by 
\eq
  \xint f^{abc}\e^{\m\n\rho\s}B^{a}_{\m\n}A^{b}_{\rho}A^{c}_{\s}  \ .
\eqn{candidate}
However, from 
\eq
s \xint f^{abc}\e^{\m\n\rho\s}B^{a}_{\m\n}A^{b}_{\rho}A^{c}_{\s} = 2 \xint
f^{abc}\e^{\m\n\rho\s}\Lp \pa_{\m} \eta^{a}_{\n}A^{b}_{\rho}A^{c}_{\s} +
B^{a}_{\m\n}\pa_{\rho}c^{b}A^{c}_{\s} \Rp \, .
\eqn{teste}
we see that the variation of \equ{candidate} unavoidably
contains $\eta^{a}_{\n}$, so that
$\lp S_{1},S_{2} \rp$ is a nontrivial element of the BRST
cohomology. 

It follows then that the only way to consistently implement the
master equation to second order in the deformation parameters is to let 
the coefficients of the nontrivial antibrackets \equ{importantes}  vanish,
{\i.e.}
\eq\ba{ll}
 g \m \,& = \,\, 0 \, , \\ \es
 g m \,& = \,\, 0 \, .
\ea\eqn{gl}
This means that, as long as we insist in adding the Yang-Mills
coupling to the free action,
then necessarily $\m = 0$ and $m = 0$. Otherwise,
we can keep the massive terms $\m $ and $m$ but
the Yang-Mills interaction is irremediably lost, and we
are left with the two massive abelian models of eqs.\equ{ab-action} and
\equ {ab-mu-action} for a set of $n$ noninteracting fields. [Both mass terms can in
fact be consistently considered simultaneously because both $(S_2, S_3)$ and
$(S_3,S_3)$ are BRST-exact].

Therefore, it is not possible to generalize the massive
action \equ{ab-action} to a nonabelian local, power counting renormalizable,
interacting theory  while preserving the same field content and the same set
of local symmetries.  This concludes the proof of the no-go 
theorem on the introduction of a topological
mass in four dimensions for a non-abelian group.  This
obstruction is in sharp contrast with the
three-dimensional Chern-Simons construction that allows
any gauge group with an invariant metric\footnote{Our 
analysis provides another instance where a duality
transformation (here between the two-form $B$ and the scalar
field of the Higgs mechanism) that can be defined in the abelian case 
cannot be implemented in the non-abelian extension \cite{Deser}.}.

\section{Comments}
\setcounter{equation}{0}

As any no-go theorem, our result is no stronger than the
assumptions underlying it.  These are two-fold.  First,
we have excluded non power-counting renormalizable couplings since the goal is
to construct an alternative to the Higgs mechanism with the
same good quantum properties.
If one relaxes this condition -- and this is suggested
by the more modern approach to renormalization advocated in
\cite{GW} --, one can construct
a Lagrangian that incorporates both the topological mass term and the 
Yang-Mills coupling.  This Lagrangian contains also the non
power-counting renormalizable Freedman-Townsend interaction and has
been written down explicitly in \cite{FT} (see also \cite{Thierry-Mieg}).
It contains two independent coupling constants, one, $m$,
with the dimensions of a mass and the other, $\alpha$,
with the dimensions of an inverse mass.  It reads
\begin{equation}
L = \frac{1}{4}\e^{\m\n\rho\s}B^a_{\m\n}\Phi^{a}_{\rho \s} -
\frac{1}{2} \b^{a \m} \b^a_{\m} -\frac{1}{2}  F^a_{\m \n} F^a_{\m \n}
\label{Full}
\end{equation}
where $\b^{a \m}$ is an auxiliary field equal on-shell to the
dual of $H^a$ in the free limit.  In \equ{Full}, $F^a_{\m \n}$
is the standard non-abelian field strength
\begin{equation}
F^a_{\m \n} = \partial_\mu A^a_\nu -  \partial_\nu A^a_\mu
+ g f^a_{bc} A^b_\mu A^c_\nu , \; \; g = m \alpha,
\end{equation}
while $\Phi^{a}_{\m \n}$ is given by a similar expression
with $\b^{a}_{\m} + m A^a_\mu$ in place of $A^a_\mu$ and
$\alpha$ in place of $g$,
\begin{equation}
\Phi^{a}_{\m \n} = \partial_\mu\b^{a}_{\n} - \partial_\n \b^{a}_{\m}
+ m(\partial_\mu A^a_\nu -  \partial_\nu A^a_\mu)
+ \alpha f^a_{bc} (\b^{b}_{\m} + m A^b_\mu)
(\b^{c}_{\n} + m A^c_\n).
\end{equation}
The first order deformations are here the topological mass
term $ m\e^{\m\n\rho\s}B^a_{\m\n}(\partial_\rho A^a_\s - 
\partial_\s A^a_\rho)$  and the Freedman-Townsend vertex
$\alpha f_{abc} \e^{\m\n\rho\s}B^a_{\m\n} \b^{b}_{\rho}\b^{c}_{\s}$.
The Yang-Mills coupling emerges
at second order, the Yang-Mills coupling constant $g$ being related
to the topological mass $m$ and the Freedman-Townsend coupling 
constant $\alpha$ through $g = m \alpha$.
The Lagrangian \equ{Full} contains terms of up to fourth
order in the coupling constants.  Of course, one could equivalently
include the mass term in the starting free lagrangian.  In that case, the
expansion in the coupling constant stops at second order (see \cite{Khoudeir}
for a first order formulation along these lines).

The second assumption underlying our analysis is that the
interacting theory possesses the same field content and the same number of
independent gauge symmetries as the free theory.  This requirement appears to
be necessary in order for perturbation theory about the quadratic piece
of the action to be
carried out.  Now, one may also give up this second assumption.  This has
been done in \cite{attempt}, where a non-abelian
generalization of the abelian action \equ{ab-action}
has been proposed, in which the abelian field strengths $F$ are
replaced by the non-abelian ones, and the derivatives
of $B$ are replaced by covariant derivatives (the
two-form $B$ transforming in the adjoint).  This generalization
is invariant under the standard Yang-Mills gauge transformations, but
not under \equ{ab-tens-tr} or any generalization
thereof since it corresponds to introducing a minimal
coupling of the two-form $B$ to the Yang-Mills connection $A$, and this was 
excluded above.  The generalization does not
provide a ``consistent deformation" of the abelian theory
because it has less gauge symmetries.  This does not mean, however, 
that it is physically inconsistent, but that the free theory
cannot be used straightforwardly as a starting point for
the standard perturbative expansion (obscuring in particular
the meaning of power-counting perturbative renormalizability).
Actually, the Hamiltonian analysis performed along the conventional
Dirac lines -- or any other equivalent method --
is rather direct and  seems to indicate that the theory is acceptable,
although there are difficulties related to the fact that some Poisson
bracket matrices have varying ranks in phase space.

One may reinstate (a non-abelian version of) the gauge
symmetry \equ{ab-tens-tr}  by adding an extra field
transforming appropriately \cite{attempt}.  But even after
this is done, the theory does not have the same number 
of gauge symmetries as the abelian limit, since in that limit, the extra
field disappears from the Lagrangian: the abelian theory with the
additional field has the additional gauge freedom of
shifting independently the extra field.  Our no-go theorem on
the absence of deformations preserving {\em all} the
gauge symmetries of the free theory applies
in fact to any equivalent formulation 
of the abelian starting point that can be obtained 
by adding extra auxiliary or pure gauge fields,
as the cohomological theorems of \cite{BBH1} (section 15)
indicate.

A construction similar to the above mass generation mechanism
has been proposed recently in three dimensions by
Jackiw and Pi \cite{Jackiw}.  What plays 
there the role of the two-form $B^a_{\m \n}$
is a vector field $B^a_\mu$.  An analog of
the above no-go theorem for the non-abelian case has been demonstrated 
long ago in \cite{Arnowitt}. 
The authors of \cite{Jackiw} 
analyse  the non-abelian theory with less gauge symmetries that
one obtains by minimally coupling the additional vector field $B^a_\mu$
to $A^a_\mu$.  The resulting theory suffers from the same difficulties
as the theory of \cite{attempt} since a standard perturbation expansion
about the free limit cannot be carried out.
Whether this theory, or the four-dimensional version of \cite{attempt},
can be quantized in a tractable and meaningful way
is therefore still an open question.  This point deserves further
investigation in view of the attractive features of
the model exhibited in \cite{Jackiw}.

\vspace{2cm}
\noindent
{\Large {\bf Acknowledgments}}

MH is grateful to Bernard Julia for pointing out to him
reference \cite{Thierry-Mieg}.  He also thanks the
Physics Departments of UERJ, UFRJ and CBPF (Rio) for kind
hospitality while this work was being carried out.  Finally,
the Conselho Nacional de Pesquisa e Desenvolvimento, CNP$q$ Brazil, is
gratefully acknowledged for financial support.



\begin{thebibliography}{19}

\bibitem{cs} R. Jackiw and S. Templeton, \pr{D23}{81}{2291} 
W. Siegel, \np{B156}{79}{135} 
J.F.  Schonfeld, \np{B185}{81}{157} 
S. Deser, R. Jackiw and S. Templeton, \annp{140}{82}{372}

\bibitem{fourd}
		E. Cremmer and J. Scherk, \np{B72}{74}{117} 
		C.R. Hagen, \pr{D19}{79}{2367} 
		T.J. Allen, M.J. Bowick and A. Lahiri,  \mpl{A6}{91}{559} 
                 R. Amorim and J. Barcelos-Neto, \mpl{A10}{95}{917}

\bibitem{def} G. Barnich and M. Henneaux, \pl{B311}{93}{123} 

\bibitem{FT}  D.\ Z.\ Freedman and P.\ K.\ Townsend, {\em Nucl.\ Phys.}\
{\bf B177} (1981) 282;

\bibitem{attempt} A. Lahiri, {\it Generating Vector Boson Masses}, 
hep-th/9301060; 
 A. Lahiri, \pr{55}{97}{5045} 
 D.S. Hwang and C. Y. Lee, \jmp{38}{97}{30}
J. Barcelos-Neto, A. Cabo and M.B.D. Silva, \zp{C72}{96}{345}
                  J. Barcelos-Neto and S. Rabello, {\it Mass generation for gauge fields in
		  the Salam-Weinberg theory without Higgs bosons}, to appear in
		  {\em Z. Phys. C}, hep-th/9601076; 

\bibitem{Julia} B. Julia, in {\em Recent Developments in Quantum Field Theory},
J. Ambjorn, B.J. Durhuus and J. L. Petersen eds, Elsevier
(1985) pp 215-225; B. Julia, in {\em Topological and Geometrical Methods in Field
Theory}, J. Hietarinta and J. Westerholm eds, World Scientific
(1986) pp325-339;

\bibitem{Stasheff} J. Stasheff, {\em Deformation Theory
and the Batalin-Vilkovisky Master Equation}, q-alg/9702012;

\bibitem{ht} M. Henneaux and C. Teitelboim, {\it Quantization of Gauge Systems}, 
Princeton University Press, Princeton, NJ, 1992; 

\bibitem{pf} M. Henneaux, {\em Phys. Lett.} {\bf 368B} (1996) 83;
M. Henneaux, B. Knaepen and C. Schomblond, \cmp{186}{97}{137}
M. Henneaux and B. Knaepen, {\it All consistent interactions for
exterior form gauge fields}, hep-th/9706119;

\bibitem{CM}  G.F. Chapline and N.S. Manton,
{\em Phys. Lett.} {\bf B120} (1983) 105;
H. Nicolai and P.K. Townsend, {\em
Phys. Lett.} {\bf 98B} (1981) 257;
A. H. Chamseddine, {\em Nucl. Phys.}
{\bf B185} (1981) 403; {\em Phys. Rev.} {\bf D24} (1981) 3065;
E. Bergshoeff, M. de Roo, B. de
Wit and P. van Nieuwenhuizen, {\em Nucl. Phys.} {\bf B195}
(1982) 97;

\bibitem{KR} M. Kalb and P. Ramond, \pr{D9}{74}{2273} 

\bibitem{DeserWitten} S. Deser and E. Witten, \np{178}{81}{491}

\bibitem{Deser} S. Deser and C. Teitelboim, {\em Phys. Rev.} {\bf D 13}
(1976) 1572;

\bibitem{GW} J.~Gomis and S.~Weinberg, {\em Nucl. Phys.\/} {\bf B469}
(1996) 473; S. Weinberg, {\it The Quantum Theory of Fields},
volumes I and II, Cambridge University Press, Cambridge 1995 and 1996;

\bibitem{Thierry-Mieg} J. Thierry-Mieg, \np{B335}{90}{334}

\bibitem{Khoudeir} A. Khoudeir, \mpl{A11}{96}{2489}

\bibitem{BBH1} G. Barnich, F. Brandt and M. Henneaux,
{\em Commun.Math.Phys.} {\bf 174} (1995) 57;

\bibitem{Jackiw} R. Jackiw and S.-Y. Pi, \pl{403B}{97}{297}

\bibitem{Arnowitt} R. Arnowitt and S. Deser, {\em Nucl. Phys.}
{\bf 49} (1963) 133.

\end{thebibliography}
\end{document}